\documentclass[12pt]{article}

\usepackage{multirow}
\usepackage{amsmath,amssymb,graphicx, graphics, color} 
\usepackage{bbm}
\usepackage{cite}
\usepackage{latexsym}
\usepackage{cancel}

\usepackage[normalem]{ulem} 

\usepackage{dsfont}

\newcommand{\drawsquare}[2]{\hbox{%
\rule{#2pt}{#1pt}\hskip-#2pt
\rule{#1pt}{#2pt}\hskip-#1pt
\rule[#1pt]{#1pt}{#2pt}}\rule[#1pt]{#2pt}{#2pt}\hskip-#2pt
\rule{#2pt}{#1pt}}
\newcommand{\Yfund}{\drawsquare{7}{0.6}}

\newcommand{\centeron}[2]{{\setbox0=\hbox{#1}\setbox1=\hbox{#2}\ifdim
\wd1>\wd0\kern.5\wd1\kern-.5\wd0\fi \copy0
\kern-.5\wd0\kern-.5\wd1\copy1\ifdim\wd0>\wd1
                                   \kern.5\wd0\kern-.5\wd1\fi}}
\newcommand{\ltap}{\>\centeron{\raise.35ex\hbox{$<$}}
                           {\lower.65ex\hbox{$\sim$}}\>}
\newcommand{\gtap}{\>\centeron{\raise.35ex\hbox{$>$}}
                           {\lower.65ex\hbox{$\sim$}}\>}
\newcommand{\gsim}{\mathrel{\gtap}}
\newcommand{\lsim}{\mathrel{\ltap}}

\newcommand\ZZ{\hbox{\zfont Z\kern-.4emZ}}
\font\zfont = cmss10 

\newcommand{\fref}[1]{Fig.\ \ref{f.#1}}
\newcommand{\eref}[1]{Eq.\ (\ref{e.#1})}

\newcommand{\sref}[1]{Section \ref{s.#1}}

\newcommand{\cref}[1]{Chapter \ref{c.#1}}

\def\kahler{K\"ahler\hspace{0.1cm}}

\newcommand{\ba}{\begin{array}}
\newcommand{\ea}{\end{array}}

\newcommand{\beq}{\begin{eqnarray}}
\newcommand{\eeq}{\end{eqnarray}}
\newcommand{\beqs}{\begin{eqnarray*}}
\newcommand{\eeqs}{\end{eqnarray*}}

\newcommand{\bal}{\begin{align}} 
\newcommand{\eal}{\end{align}}

\def\bi{\begin{itemize}}
\def\ei{\end{itemize}}
\def\ben{\begin{enumerate}}
\def\een{\end{enumerate}}
\def\bc{\begin{center}}
\def\ec{\end{center}}
\def\bt{\begin{table}}
\def\et{\end{table}}
\def\btb{\begin{tabular}}
\def\etb{\end{tabular}}

\def\mass2{mass${}^2$}

\def\csaki{Cs\'aki}

\begin{document}
\bibliographystyle{unsrt}
\begin{titlepage}

\vskip2.5cm
\begin{center}
\vspace*{5mm}
{\huge \bf Supersymmetry Breaking Triggered by Monopoles}
\end{center}
\vskip0.2cm

\begin{center}
{\bf Csaba \csaki$^1$, David Curtin$^{1,2}$, Vikram Rentala$^{3,4}$,\vspace{2mm} \\ Yuri Shirman$^4$, John Terning}$^5$

\end{center}
\vskip 8pt

\begin{center}
{\it $^1$Department of Physics, LEPP, Cornell University, Ithaca, NY 14853.} \\
\vspace*{0.3cm}
{\it $^2$ C. N. Yang Institute for Theoretical Physics\\ Stony Brook University, Stony Brook, NY 11794.}\\
\vspace*{0.3cm}
{\it $^3$ Department of Physics, University of Arizona, Tucson, AZ 85721.}\\
\vspace*{0.3cm}
{\it $^4$ Department of Physics, University of California, Irvine, CA 92697.}\\
\vspace*{0.3cm}
{\it $^5$ Department of Physics, University of California, Davis, CA 95616.}\\
\vspace*{0.3cm}

\vspace*{0.1cm}

{\tt csaki@cornell.edu,
curtin@insti.physics.sunysb.edu,
vikram.rentala@gmail.com,
yshirman@uci.edu, terning@physics.ucdavis.edu}
\end{center}

\vglue 0.3truecm

\begin{abstract}
We investigate ${\mathcal N}=1$ supersymmetric gauge theories  where monopole condensation triggers
supersymmetry breaking in a metastable vacuum.  The low-energy effective theory
is an O'Raifeartaigh-like model of the kind investigated recently by Shih where the $R$-symmetry can be spontaneously broken.
We examine several implementations with varying degrees of phenomenological interest.

\end{abstract}

\end{titlepage}

\section{Introduction}
\label{s.intro} \setcounter{equation}{0} \setcounter{footnote}{0}

Magnetic monopoles are fascinating for many reasons. Decades before Grand Unified Theories were considered, Dirac \cite{DiracChargeQuant} realized that the  existence of monopoles implies charge quantization. It turns out that monopoles and GUTs are intrinsically connected, and monopoles can arise dynamically as topologically stable gauge field configurations from spontaneous gauge symmetry breaking \cite{topologicalMonopoles}. Their dynamics appear quite distinct from other kinds of objects in quantum field theory, and actions that incorporate monopoles into a theory with electric charges have to be either Lorentz-violating \cite{DiracLagrangian} or non-local \cite{ZwanzigerLagrangian}. A magnetically charged condensate leads to a magnetic dual Meissner effect and represents one possible explanation for confinement \cite{condensationconfinement}. (For an excellent review of these ideas the reader is directed to \cite{Preskill}.)

In 1994 Seiberg and Witten \cite{SeibergWitten} were able to use elliptic curves to find the low-energy effective action of $\mathcal{N} = 2$ supersymmetric $SU(2)$ gauge theories. As the gauge symmetry is broken to $U(1)$  we would expect to find heavy topological monopoles and dyons. They discovered that some of those topological states become massless weakly coupled particles at certain singular points on the moduli space, where the electric gauge coupling diverges. Furthermore, softly breaking these theories to $\mathcal{N}=1$ supersymmetry (SUSY) lifted the moduli space and induced the massless monopole to condense, leading to electric confinement and providing an illuminating perspective on the well-known $\mathcal{N}=1$ phenomenon of gaugino condensation. (A pedagogical introduction can be found in \cite{SWreview}.) These results were soon generalized to higher gauge groups \cite{SWgeneralizations}. The higher-dimensional moduli spaces of these theories contain singular submanifolds where both electric and magnetic charges of the same $U(1)$ gauge group become simultaneously massless, providing the first example of a self-consistent quantum field theory where this particle content arises dynamically. It is also possible to apply these methods to the analysis of minimally supersymmetric $\mathcal{N} = 1$ theories in the Coulomb phase and extract the holomorphic parts of the low-energy effective action \cite{SWforN=1}.

SUSY is, of course, extremely interesting for phenomenological reasons, the most important one being the stabilization of the weak scale. While there are several possible mechanisms for breaking supersymmetry \cite{SUSUbreakingpossibilities} and mediating its breaking to the supersymmetric standard model, no clear favorite has emerged. It is therefore prudent to continue looking for new ways of breaking SUSY. The unique properties of monopoles, and the fact that they arise as light states dynamically and calculably in some theories, motivate the construction of SUSY-breaking models that rely on monopole dynamics. The hope is that eventually some new mechanisms with desirable, novel features might be found. In this paper, we initiate the study of such constructions\footnote{
Theories with light monopoles and dyons are also worth studying since they might yield a possible mechanism of electroweak symmetry breaking \cite{EWSBbymonopoles}.}.

Models involving meta-stable SUSY-breaking \cite{olderMetaStable} circumvent the restrictive Witten index constraint \cite{WittenIndex}. The topic enjoyed a fresh surge of interest since Intriligator, Seiberg, and Shih showed how theories as simple as SQCD in the free magnetic phase can feature metastable SUSY-breaking vacua \cite{ISS}, and since then many models have been proposed to incorporate these ideas into a phenomenologically viable model \cite{ISSmodels}.   The authors of \cite{ISS} also suggested that  $\mathcal{N}=2$ SYM might plausibly feature such SUSY-breaking vacua, and their intuition turned out to be correct. Deformed $\mathcal{N}=2$ theories  can generate SUSY-breaking local minima  at generic points of their moduli spaces \cite{OOP}, but the metastable vacua do not lie on the singularity and the monopoles of the theory play no role. Ref. \cite{AMOS} considers an $\mathcal{N}=2$ model in the Coulomb phase softly broken to $\mathcal{N}=1$ which classically breaks SUSY in a metastable vacuum via a Fayet-Illiopoulos term. They find that the SUSY-breaking survives the nonperturbative quantum effects, and that light dyons undergo condensation at the meta-stable SUSY-breaking minimum.

Here, we want to construct a dynamical model of supersymmetry-breaking where SUSY would be restored in the absence of a monopole condensate.   Our starting point will be the $\mathcal{N}=1$  $SU(2)^3$ model \cite{CEFS}, where the gauge symmetry is broken to $U(1)$ on the moduli space and there are singular submanifolds on which monopoles or dyons become massless. We then deform the theory to obtain monopole condensation near a point on the singular submanifold of the moduli space. In the low-energy limit we find an effective O'Raifeartaigh-type model of the form recently investigated by Shih \cite{ShihModel}  which features a metastable SUSY-breaking minimum.

The paper is structured as follows. In \sref{CEFS} we briefly review the $\mathcal{N} = 1$ $SU(2)^3$ model with massless monopoles and parameterize our ignorance of the K\"ahler potential to write down an effective theory near a point on the singular submanifold. In \sref{shih} we review the Shih-O'Raifeartaigh model and derive the scaling behavior of some important quantities determined at 1-loop. We then deform the $SU(2)^3$ model to let monopole condensation trigger supersymmetry breaking in \sref{model1}. \sref{model2and3} explores some variations of this model, and we conclude with \sref{conclusion}.

\section{The $SU(2)^3$ Model}
\label{s.CEFS} \setcounter{equation}{0} \setcounter{footnote}{0}

The basis for our model of SUSY-breaking is the $SU(2)^3$ model \cite{CEFS}. After briefly reviewing its main features, we will expand the theory around a particular point in moduli space in order to explicitly parameterize our ignorance of the incalculable \kahler potential.

\subsection{Review}

Our starting point is an ${\mathcal N}= 1$ SUSY model  \cite{CEFS} with a $SU(2)_1 \times SU(2)_2 \times SU(2)_3$ gauge symmetry that is is broken down to a diagonal $U(1)$  at low energies. This makes it possible to apply Seiberg-Witten methods \cite{SeibergWitten, SWforN=1} to obtain information about the holomorphic quantities (the superpotential and  gauge kinetic term) of the model. The particle content of the underlying electric theory is 
\begin{equation}
\begin{array}{c|ccc}
&SU(2)_1 & SU(2)_2 & SU(2)_3\\\hline
Q_1 & \Yfund & \Yfund & 1\\
Q_2 & 1 & \Yfund & \Yfund\\
Q_3 & \Yfund & 1 & \Yfund
\end{array}
\end{equation}
The three $SU(2)_i$ gauge groups become strongly coupled below scales $\Lambda_i$. For simplicity we let $\Lambda_i = \Lambda$. The moduli space is spanned by four gauge invariants
\begin{eqnarray}
\nonumber M_i &=& \det Q_i = \frac{1}{2} (Q_i)_{\alpha \beta} (Q_i)_{\gamma \delta} \epsilon^{\alpha \gamma} \epsilon^{\beta \delta},\\
\label{e.CEFSmoduli}T &=& \frac{1}{2} (Q_1)_{\beta_1 \alpha_2} (Q_2)_{\beta_2 \alpha_3} (Q_3)_{\beta_3 \alpha_1} \epsilon^{\alpha_1 \beta_1} \epsilon^{\alpha_2 \beta_2} \epsilon^{\alpha_3 \beta_3},
\end{eqnarray}
and at generic points in the moduli space the $SU(2)^3$ gauge group is broken down to the diagonal $U(1)$, so the theory is in the Coulomb Phase. The holomorphic quantities of the low-energy theory are described by  an elliptic curve
\begin{equation}
\label{e.CEFScurve}
y^2 = \left[ x^2 - \left(\Lambda^4 M_2 + \Lambda^4 M_3 + \Lambda^4 M_1 - M_1 M_2 M_3 + T^2\right) \right]^2 - 4 \Lambda^{12}.
\end{equation}
Rescaling the variables by defining
\begin{eqnarray*}
u_{SW} &=&  2 \left( \Lambda^4 M_2 + \Lambda^4 M_3 + \Lambda^4 M_1 - M_1 M_2 M_3 + T^2  \right),\\
\Lambda_{SW}^2 &=& 2 \ \Lambda^6,
\end{eqnarray*}
we see that \eref{CEFScurve} is identical to the  ${\mathcal N}=2$ $SU(2) $ SYM curve \cite{SeibergWitten},
\begin{equation}
y^2 = \left(x^2 - \frac{1}{2} u_{SW}\right)^2 - \Lambda_{SW}^4~.
\end{equation}

The elliptic curve represents a torus with complex structure, and the low-energy $U(1)_{eff}$ holomorphic gauge coupling is given as the ratio of the two periods of the torus.
The torus can be transformed by an $SL(2, \mathbb Z)$ transformation, which corresponds to transforming the low-energy effective $U(1)$ gauge theory into a dyonic dual description. In the electric description the electric gauge coupling approaches zero as $u \rightarrow \infty$. The roots of the ${\mathcal N}= 2$ $SU(2)$ SYM elliptic curve are degenerate for $u = \pm 2 \Lambda^2$, meaning that the torus becomes singular on the corresponding submanifolds of the full moduli space. This causes the electric gauge coupling to diverge on these submanifolds, whereas the magnetic/dyonic gauge coupling goes to zero. Therefore certain monopoles or dyons, which are large, massive and strongly coupled topological objects in the weakly coupled electric regime $u \rightarrow \infty $, become elementary, light, and weakly coupled (the magnetic coupling goes to zero) near the respective singularities. The monopoles and dyons of the $SU(2)^3$ model become massless when
\begin{equation}
\label{e.singularitycondition}
 \Lambda^4 M_2 + \Lambda^4 M_3 + \Lambda^4 M_1 - M_1 M_2 M_3 + T^2  \pm 2 \Lambda^6 = 0.
\end{equation}

Near these two points in moduli space, the effective potential can be approximated as
\begin{eqnarray}
\nonumber W_{eff} &= &  \frac{1}{\Lambda^5}\left[ - \Lambda^4 M_1 -  \Lambda^4 M_2 - \Lambda^4 M_3 + M_1 M_2 M_3 -  T^2  \pm 2 \Lambda^6 \right] E_\pm \tilde E_\pm  + \{\mathrm{HOT}\},
\end{eqnarray}
where $E_\pm, \tilde E_\pm$ are monopoles/dyons, which are light, elementary and weakly coupled near the singularity. 
The higher-order terms \{HOT\} only contain higher powers of the term in square brackets and cannot change the location of the singularity. 
Higher powers of monopoles/dyons in the superpotential are forbidden by global symmetries (including an anomalous $U(1)_R$) and holomorphy. Rescaling the moduli to have mass dimension 1, this becomes 
\begin{equation}
\label{e.CEFSsuperpot}
W_{eff} = \left[ - M_1 - M_2 - M_3 + \frac{M_1 M_2 M_3}{\Lambda^2} - \frac{T^2}{\Lambda}  \pm 2 \Lambda \right] E_\pm \tilde E_\pm + \{\mathrm{HOT}\}~.
\end{equation}

\subsection{Effective theory near a singular point}

We will deform the $SU(2)^3$ model in a way that will lead to SUSY-breaking
triggered by monopole condensation. To that end we want to find the effective theory near a singular point on the moduli space, defined by 
\begin{equation}
\label{e.pointinMS}
M_1 = 2 \Lambda \ \ , \ \ \ \ \ \ \ M_{2,3} = 0 \ \ , \ \ \ \ \ T = 0.
\end{equation}
The existence and stability of a SUSY-breaking minimum (after perturbations are included) in the vicinity of this
point will depend on the exact form of the \kahler potential, which cannot be calculated using Seiberg-Witten techniques in an $\mathcal{N} = 1$ theory. Instead we will
expand the effective Lagrangian in small fluctuations around the supersymmetric state (\ref{e.pointinMS}) and
restrict the form of the Kahler potential using unbroken global symmetries. Expanding $M_1 = 2 \Lambda + \delta M_1$, \eref{CEFSsuperpot} becomes
\begin{equation}
\label{e.CEFSsuperpot2}
W_{eff} = \left[ - \frac{T^2}{\Lambda} - \delta M_1 - M_2 - M_3  \right] E_\pm \tilde E_\pm+  \{\mathrm{HOT}\}~.
\end{equation}
where \{HOT\} now includes terms like $M^2E_\pm \tilde E_\pm/\Lambda$ and  $T^2 M E_\pm \tilde E_\pm/\Lambda^2$. 
(We explictly keep the $T^2/\Lambda$ term because it gives the lowest-order contribution to the potential for $T$.)
While the \kahler potential is not determined by holomorphy, the weakly coupled degrees of freedom near the singular point are the monopoles and the moduli, and the \kahler potential is non-singular in terms of these fields with an expansion in inverse powers of $\Lambda$.
The global symmetries are then used to constrain the \kahler potential. There is an $S_3$ symmetry which switches the $M_i$'s and $SU(2)_i$'s around, as well as a slightly less obvious $\mathbb{Z}_4$ which acts on each of the electric quarks as
\begin{equation}
Q_i \rightarrow e^{i n \pi/2} Q_i.
\end{equation}
This is an anomaly-free $\mathbb{Z}_4$ subgroup \cite{discretesymmetries} of anomalous $U(1)$ global symmetry under which each of the electric quarks has charge $1$. Under this symmetry, the moduli transform as $M_i \rightarrow e^{i n \pi} M_i$ and $T \rightarrow e^{i n \pi/2} T$,
 meaning that $M_i$ and $T^2$ both have charge 2 under the $\mathbb{Z}_4$. The $x$ and $y$ coordinates of the elliptic curve, \eref{CEFScurve}, have charge 2 and 0 respectively, while $\Lambda$ has charge 0.  
 
Around the point in moduli space  (\ref{e.pointinMS}), the global symmetries of the model are broken from $S_3$ to $S_2$, which exchanges $M_2$ and $M_3$, and $\mathbb{Z}_4$ is broken to $\mathbb{Z}_2$, under which $T \rightarrow - T$ and the $M_i$ are singlets. 
Defining a field basis  $ \varphi^i =  (\delta M_1, M_2, M_3, T, E_+, \tilde E_+)$, we write the \kahler potential as an expansion in the small fluctuations 
\begin{equation}
K =  \varphi^\dagger_j K^j_i \varphi^i + \mathcal{O}\left(\frac{\varphi^3}{\Lambda}\right),
\end{equation}
where $K^j_i$ is a hermitian positive-definite matrix. The symmetries then restrict $K^j_i$ to be of the form
\begin{equation}
K^j_i = \left[ \begin{array}{cccccc}
\alpha & \delta e^{i \theta}& \delta e^{i \theta} & 0 & 0 & 0 \\
\delta e^{-i \theta} & \beta & \gamma & 0 & 0  & 0 \\
\delta e^{-i \theta} & \gamma & \beta & 0 & 0 & 0 \\
0 & 0 & 0 & \kappa & 0 & 0\\
0 & 0 & 0 & 0 & \eta & 0\\
0 & 0 & 0 & 0 & 0 & \eta
\end{array} \right],
\end{equation}
where all parameters are real and positive definiteness requires $\kappa > 0$, $\eta > 0$, $\beta > \gamma$, $\alpha ( \beta + \gamma) > 2 \delta^2$. The precise values of these parameters are unknown but presumably $\mathcal{O}(1)$.

We can now define new degrees of freedom $\tilde M_i$ to diagonalize the upper $3 \times 3$ corner of $K^j_i$. Upon rescaling, all degrees of freedom can then be made canonical to quadratic order in the \kahler potential, giving a effective superpotential valid in the neighborhood of \eref{pointinMS},
\begin{equation}
\label{e.CEFSsuperpot1}
W_{eff} = \left[ a \tilde M_1 + b \tilde M_2 + c \tilde M_3 - d \frac{T^2}{\Lambda}\right] E_+ \tilde E_+,
\end{equation}
where the coefficients $a,b,c,d$ are unknown complex $\mathcal{O}(1)$ numbers into which we have absorbed the canonical rescaling of the monopole fields. As long as  the $S_2$ symmetry is unbroken one can show that $c = 0$, but we include this coefficient for generality since it will be induced perturbatively by explicit $S_2$ breaking effects, as discussed in Section \ref{s.model1}.

\section{The Shih-O'Raifeartaigh Model}
\label{s.shih} \setcounter{equation}{0} \setcounter{footnote}{0}

Triggering SUSY-breaking via monopole condensation can be achieved by deforming the $SU(2)^3$ model to resemble the Shih-O'Raifeartaigh model \cite{ShihModel} in the low-energy limit (near a singular point of moduli space). In this section we will briefly review the Shih-O'Raifeartaigh model and then derive some scaling behavior which will be important in ensuring the stability of our SUSY-breaking local minimum against incalculable corrections.

\subsection{The Model}

In \cite{ShihModel}, Shih wrote down an O'Raifeartaigh model with a single pseudomodulus and $R$-charges other than 0 or 2 which can break $R$-symmetry spontaneously without tuning.\footnote{See \cite{genORmorePM} for some studies of spontaneous $R$-breaking in models with multiple pseudomoduli.} The superpotential is
\begin{equation}
W = \lambda X \phi_1 \phi_2 + m_1 \phi_1 \phi_3 + \frac{1}{2} m_2 \phi_2^2 + f X~.
\end{equation}
The $R$-charges are $R_X = 2, R_{\phi_1} = -1, R_{\phi_2} = 1, R_{\phi_3} = 3$. The tree-level scalar potential is
\begin{equation}
V = |\lambda \phi_1 \phi_2 + f|^2 + |\lambda X \phi_2 + m_1 \phi_3|^2 + |\lambda X \phi_1 + m_2 \phi_2|^2 + |m_1 \phi_1|^2~.
\end{equation}
Via field redefinitions we can let all the parameters be real and positive. It is useful to define the two dimensionless parameters
\begin{equation}
y = \frac{\lambda f}{m_1 m_2}, \ \ \ \ \ \ \ r = \frac{m_2}{m_1}.
\end{equation}
For $y < 1$, there exists a pseudoflat direction that breaks SUSY:
\begin{equation}
\phi_i = 0, \ \ \ \ \ \ \ X \  \mathrm{arbitrary}\ \ \ \ \Rightarrow \ \ \ \ \ \langle V \rangle = f^2~.
\end{equation}
The field $X$ is a pseudomodulus, meaning it does not receive a potential at tree-level but does get one at 1-loop. The pseudomoduli space is stable in a neighborhood of the origin $|X| < X_{max}$, where \begin{equation}
\label{e.Xmax}
X_{max} \equiv \frac{m_1}{\lambda} \frac{1-y^2}{2 y}~.
\end{equation}
The 1-loop Coleman-Weinberg potential \cite{Coleman} can stabilize $X$ at the origin for $r \lsim 2$ or break the $R$-symmetry and induce a non-zero $\langle X\rangle$ for $r \gsim 2$, see \fref{shihVcw}. There is also a SUSY-runaway:
\begin{equation}
\label{e.shihrunaway}
\phi_1 = \sqrt{\frac{f m_2}{\lambda^2 X}} \ \ , \ \ \ \ \ 
\phi_2 = - \sqrt{\frac{f X}{m_2}} \ \ , \ \ \ \ \ 
\phi_3 = \sqrt{\frac{\lambda^2 f X^3}{m_1^2 m_2}} \ \ , \ \ \ \ \ 
X \rightarrow \infty
\end{equation}
Along this runaway direction, $V = f m_1^2 m_2/(\lambda^2 |X|)$, so the value of $|X|$ at which the potential energy becomes equal to the false vacuum energy $f^2$ is
\begin{equation}
|X| = X_{cross} \equiv \frac{m_1}{\lambda y} = \sqrt{\frac{f}{\lambda r y^3}}~.
\end{equation}
For smaller values of $|X|$, the potential energy is larger than $f^2$ along the runaway direction. The width of the potential barrier that separates the false vacuum from the runaway scales with negative powers of $y$ and $\lambda$, so if either is small the parametric longevity of the SUSY-breaking minimum can be guaranteed.

\begin{figure}
\begin{center}
\begin{tabular}{cc}
\includegraphics[width=7cm]{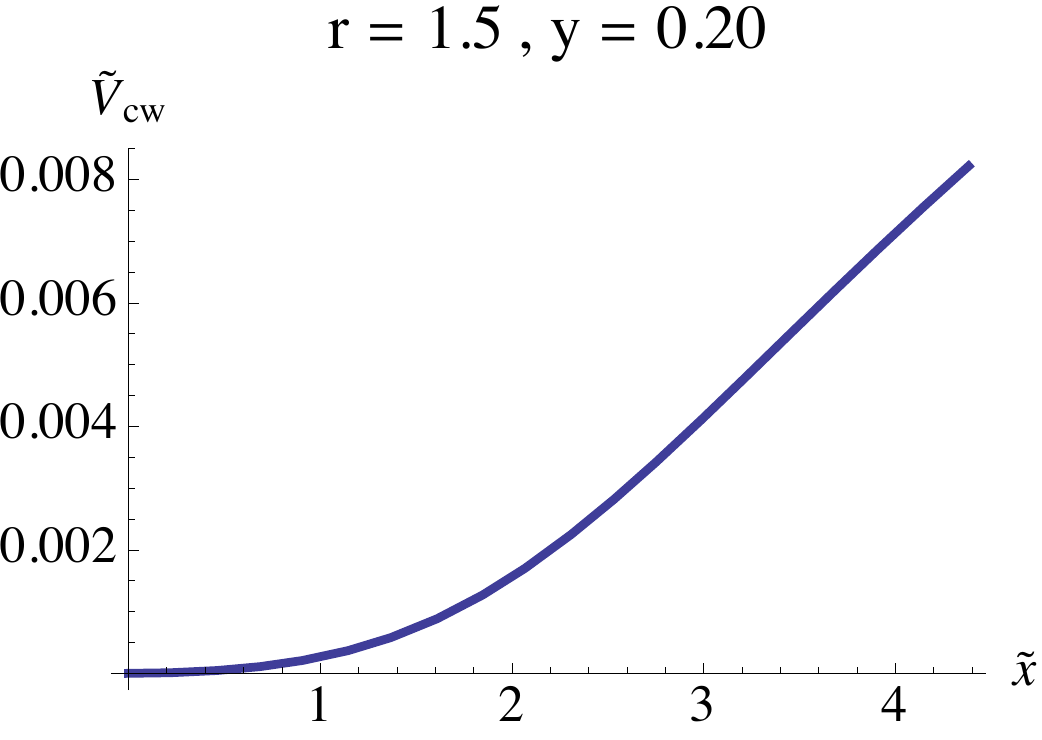}
&
\includegraphics[width=7cm]{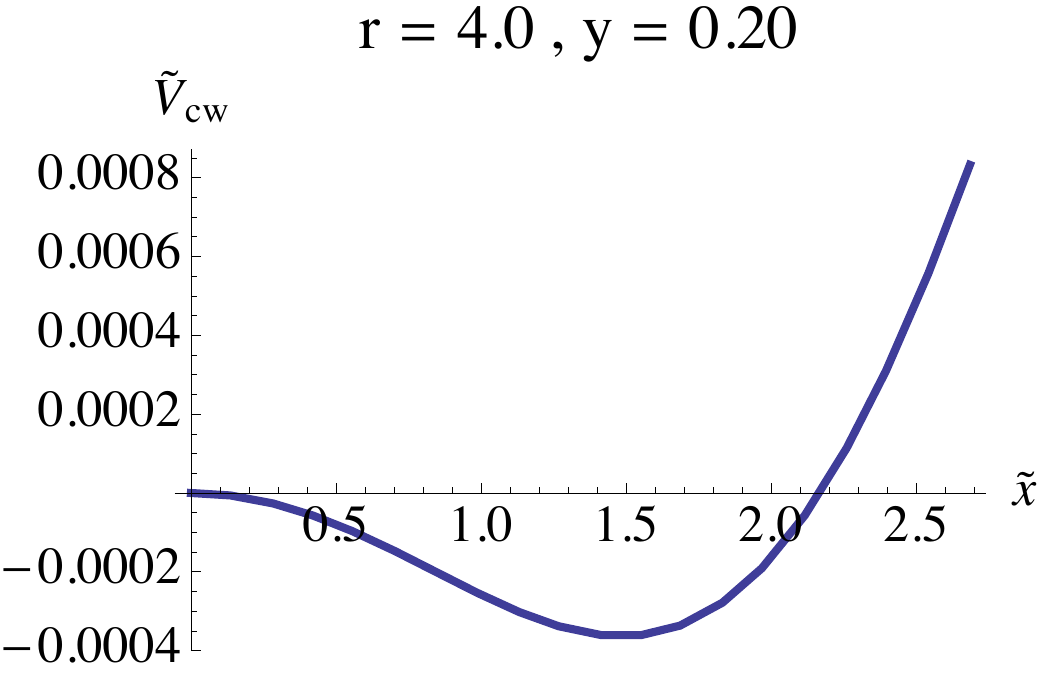}
\end{tabular} \vspace*{-0mm}
\end{center}
\caption{Two examples of the 1-loop Coleman-Weinberg potential $V_\mathrm{CW}$ (shifted by a constant) that generate zero and nonzero VEV for $X$. Note that $|X| = \sqrt{\frac{f}{\lambda}} \ \tilde x$ and $V_\mathrm{CW} = \lambda^2 f^2 \tilde V_\mathrm{CW}$.}
\label{f.shihVcw}
\end{figure}

\subsection{Scaling Behavior}

We will eventually construct a model which reduces to the Shih-O'Raifeartaigh model in a low-energy limit, up to incalculable \kahler corrections and higher-order terms in the superpotential. To ensure that these incalculable contributions to the scalar potential do not destabilize the false vacuum we will have to understand the scaling behavior of the pseudomodulus mass, VEV and the gradient of the potential barrier. 

The first step is to separate out the $f/\lambda$ scaling from the $r,y$ scaling by redefining a dimensionless version of  the 1-loop Coleman-Weinberg potential
\begin{equation}
V_{\mathrm{CW}}(|X|) = \frac{1}{64 \pi^2} \mathrm{Tr}(-1)^F \mathcal{M}^4 \log{\frac{\mathcal{M}^2}{\Lambda^2}}~
\end{equation}
in the following fashion:
\begin{equation}
\tilde x = \frac{ \sqrt{\lambda} |X|}{\sqrt{f}}, \ \ \ \  \tilde V_\mathrm{CW}(\tilde x) \equiv \frac{1}{\lambda^2 f^2} V_\mathrm{CW}\left(\frac{\tilde x}{\sqrt{\lambda}} \sqrt{f}\right).
\end{equation}
Then $\tilde V_\mathrm{CW}(\tilde x)$ depends only on $\tilde x, y, r$ (up to an additive constant). In these units, $X_{max}$ from \eref{Xmax} becomes
\begin{equation}
\tilde x_{max} = \frac{1-y^2}{2 \ y^{3/2} \ r^{1/2}}.
\end{equation}
We can now easily explore the $r,y$ scaling of $\tilde V_\mathrm{CW}(\tilde x)$ numerically. 

\begin{figure}
\begin{center}
\begin{tabular}{cc}
\includegraphics[width=8cm]{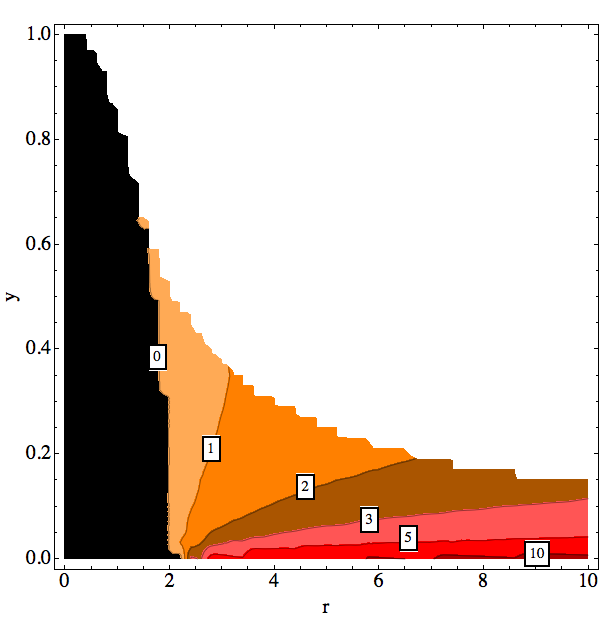}
&
\includegraphics[width=8cm]{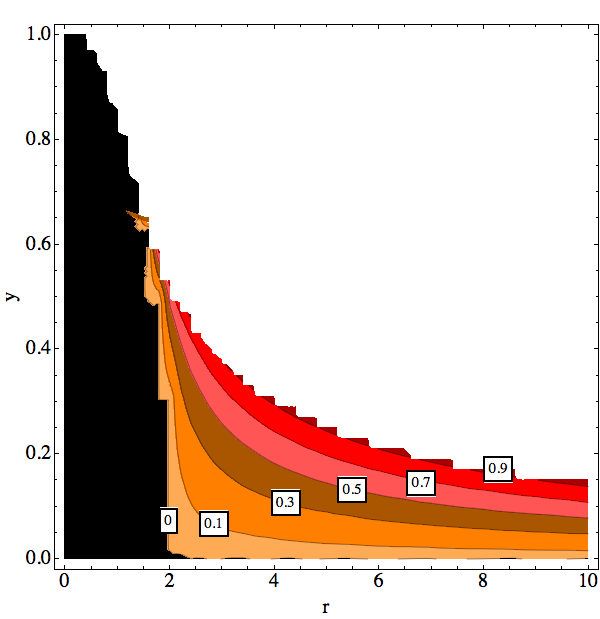}\\ (a) & (b)
\end{tabular}
\end{center} \vspace*{-5mm}
\caption{The pseudomodulus VEV $\langle |X| \rangle$ in units of (a) $\sqrt{f/\lambda}$ and (b) $X_{max}$ .  White areas indicate that the 1-loop Coleman-Weinberg potential slopes away from the origin without any local minima. Notice that for $r \gsim 2$, $R$-symmetry is spontaneously broken.}
\label{f.shihxvev}
\end{figure}

\begin{figure}
\begin{center}
\includegraphics[width=8cm]{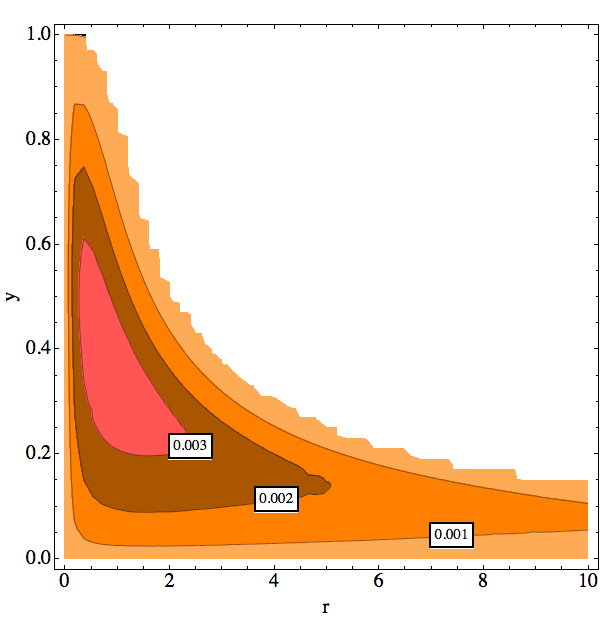}
\end{center}
\caption{The maximum value of the gradient $\frac{dV}{dX}$ in the interval $|X| \in (\langle |X| \rangle, X_{max})$,  in units of  $\sqrt{\lambda^5 f^3}$.}
\label{f.shihVcwgradient}
\end{figure}

\begin{figure}
\begin{center}
\includegraphics[width=8cm]{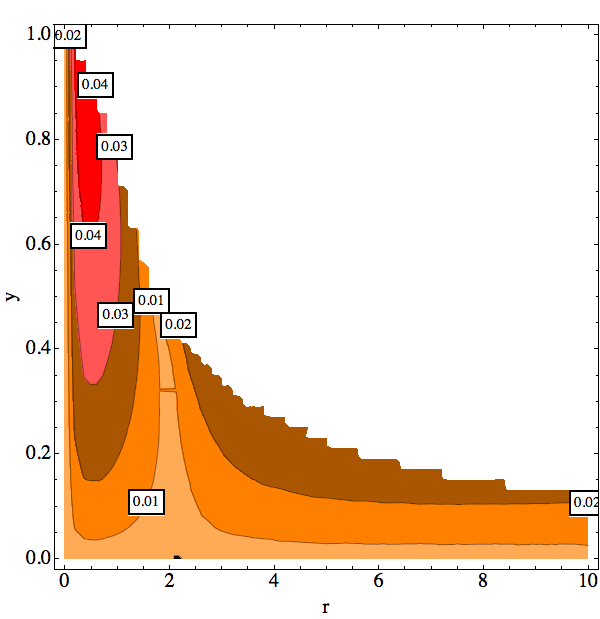}
\end{center}
\caption{The pseudomodulus mass,  $m_X$, generated by $V_\mathrm{CW}$ in units of $\sqrt{ \lambda^3 f}$.}
\label{f.shihxmass}
\end{figure}

There are two regimes of interest. For the first numerical scan we let $r \in (0,10)$ to explore the interesting $r \sim \mathcal{O}(1)$ behavior of $V_\mathrm{CW}$ and make the plots of $\langle X \rangle$, the gradient of the potential barrier and the mass of the pseudomodulus shown in Figures \ref{f.shihxvev}, \ref{f.shihVcwgradient} and \ref{f.shihxmass}. 

The second scan looked at $\log_{10}(r) \in (1,8)$ to extract scaling behaviors with $r$ and $y$ varying over many orders of magnitude. These extracted scalings turned out to give reasonable order-of-magnitude estimates for $r \sim \mathcal{O}(1)$ as well. The results can be summarized as follows:
\begin{itemize}
\item For any $r$, there exists a $y_{max} < 1$ such that one can find a local minimum $0 \leq \langle |X| \rangle < X_{max}$ on the pseudomoduli space that is stabilized by quantum corrections along the $X$-direction. In other words, \emph{$y < y_{max}$ is a requirement for metastable SUSY-breaking}, which is stronger than $y < 1$.  From the scan, we are able to extract the following scaling behavior:
\begin{equation}
\label{e.ymax}
y_{max} \approx \frac{2}{r} 
\end{equation}
where the error is $\sim 100\%$, $\sim 30\%$, and $< 1\%$ for $r \sim 1$, $\sim 10$, and $> 100$ respectively. Hence we can extract an interesting constraint for large $r$ that must be satisfied to guarantee the existence of a SUSY-breaking local minimum:
\begin{equation}
\lambda f  \lsim m_1^2 \ \ \ \ \ \ \ \ \ \ \ \ \ \ \mbox{for $r \gsim 10$}.
\end{equation}

\item The pseudomodulus VEV $\langle X \rangle$ is 0 for $r \lsim 2$. For $r \gsim 2$ a good approximation is 
\begin{equation}
\label{e.shihXvev}
\langle X \rangle \ \approx \ \frac{r y}{2} X_{max} \ = \  (1-y^2)  \ \frac{m_2}{4 \lambda} 
\end{equation}
with errors $\mathcal{O}(10\%)$, $\mathcal{O}(1\%)$ when $r \sim \mathcal{O}(1)$ and $r > 10$ respectively.

\item The maximum (positive) gradient between $\langle X \rangle$ and $X_{max}$ is roughly given by 
\begin{equation}
\label{e.shihbarriergradient}
\left[\frac{\Delta V}{\Delta |X|}\right]_{max} \ \sim \  \ \frac{1}{8 \pi^2} \sqrt{\frac{y}{r}} \ \times  \sqrt{\lambda^5 f^3} \  = \ \frac{1}{8 \pi^2} \ \frac{\lambda^3 f^2}{m_2},
\end{equation}
as long as $y$ is not very close to $y_{max}$, in which case the gradient approaches 0.

\item As shown in \fref{shihxmass}, the behavior of the pseudomodulus mass mostly depends on $y$ with the exception of the dip near $r \approx 2$.  Ignoring the dip, the mass scales as
\begin{equation}
\label{e.shihmodelpmmass}
m_X \sim \frac{1}{3} \  y \ \times \ \sqrt{\lambda^3 f} \ \ = \ \  \frac{1}{3} \ \frac{\sqrt{\lambda^5 f^3}}{m_1 m_2}
\end{equation}

\end{itemize}

Let us reexamine the lifetime of the SUSY-breaking minimum in light of these scalings. 
The pseudomodulus VEV is at $\langle X \rangle = 0$ or $\langle X \rangle \approx (1-y^2) m_2 (4 \lambda)^{-1}$.  The smallest value of $|X|$ at which the SUSY-runaway has a lower potential energy than the potential of the pseudoflat direction is $X_\mathrm{cross} = m_1 (\lambda y)^{-1}$. If $\langle X \rangle$ is not zero, one can show using $y < y_{max} \sim 2/r$ that $\langle X \rangle/X_{cross} \approx r y (1-y^2)/4 \lsim (1-y^2)/2 < 1/2$, so regardless of $r$ the barrier width scales as $X_{cross}$, or $\sim \mathcal{O}(y^{-1}\lambda^{-1})$. Therefore, if either $y$ or $\lambda$ is small, the longevity of the supersymmetry breaking vacuum is guaranteed. Since $y^{-1} \gsim r$ for a SUSY-vacuum, the presence of such vacuum when $r$ is large also guarantees its longevity.

\section{Breaking SUSY by Monopole Condensation}
\label{s.model1} \setcounter{equation}{0} \setcounter{footnote}{0}

Now we want to deform the $SU(2)^3$ model such that, near the monopole singularity, the monopoles condense and the low-energy effective theory (below the condensation scale) resembles the Shih-O'Raifeartaigh model of metastable SUSY-breaking. This mechanism should be dynamical in the sense that SUSY is restored in the weak-coupling limit $\Lambda \rightarrow 0$. To achieve this, we introduce $SU(2)^3$-singlet fields $\phi_{1,2,3}, Z, Y$ and the following tree-level superpotential to the electric theory:
\begin{eqnarray}
\nonumber W_\mathrm{tree} &=& \tilde m {(QQ)_A} + \frac{\tilde \lambda}{\Lambda_{UV}} { (QQ)_B} \phi_1 \phi_2 + \frac{1}{2} m_2 \phi_2^2 + m_1 \phi_1 \phi_3\\
&&+ \frac{a_Z}{\Lambda_{UV}} Q_1 Q_2 Q_3 Z + a_Y { (QQ)_C} Y.
\label{e.model1Wtree}
\end{eqnarray}
Here, $(QQ)_{A,B,C}$ are linear combinations of $Q_1^2, Q_2^2, Q_3^2$ and $\Lambda^2$ that become the canonical $\tilde M_{1,2,3}$ perturbations around the point, \eref{pointinMS}, in the IR. The electric quark mass $\tilde m$ and the UV-physics scale $\Lambda_{UV}$ must be much smaller and larger, respectively, than $\Lambda$ to protect the nonperturbative $SU(2)^3$ dynamics. For the same reason the Yukawa coupling $a_Y$ must be much less than unity.  The couplings $\tilde \lambda$ and $a_Z$ are perturbative. These deformations explicitly break the $S_3$ symmetry, while the $\mathbb{Z}_4$ symmetry is reduced to $\mathbb{Z}_{2}$, with $T$ and $Z$ both having charge 1. Crucially, this is still sufficient to prevent any quadratic \kahler mixing in the low-energy theory except amongst the $M_{i}$ perturbations, but in \eref{CEFSsuperpot1} a nonzero $c \sim \mathcal{O}(\mathrm{max}\{ \frac{1}{16 \pi^2} \tilde \lambda \frac{\Lambda}{\Lambda_{UV}}, \frac{1}{16 \pi^2} a_Z \frac{\Lambda}{\Lambda_{UV}}, \frac{1}{16 \pi^2} a_Y\})$ will be generated.

In the magnetic theory near the monopole singularity $W_{tree}$ is mapped to 
\begin{eqnarray}
\label{e.model1deformations}
\nonumber \delta W &=& - \mu^2 \tilde M_1 + \lambda \tilde M_2 \phi_1 \phi_2 + \frac{m_2}{2} \phi_2^2 + m_1 \phi_1 \phi_3\\
\label{e.CEFSdeformation1} && + m_Z Z T + m_Y \tilde M_3 Y,
\end{eqnarray}
where $\mu^2 \sim  m \Lambda \ll \Lambda^2$, $\Lambda \sim \tilde \lambda \ \Lambda/\Lambda_{UV} \ll 1$, $m_Z \sim a_Z \ \Lambda^2/\Lambda_{UV} \ll \Lambda$, and $m_Y \sim a_Y \Lambda \ll \Lambda$. 
By absorbing phases into the fields appropriately, all the parameters can be made real and positive.

\subsection{Meta-stable SUSY-breaking vacuum}
The rationale behind choosing the particular form of the deformations \eref{model1deformations} is the following: the mass terms for $Z T$ and $\tilde M_3 Y$  stabilize the respective moduli at the origin, while $F_{\tilde M_1} = a E_+ \tilde E_+ - \mu^2$ forces the monopoles to condense, which creates an effective tadpole for $\tilde M_2$. This generates an effective Shih-O'Raifeartaigh model, where the pseudomodulus is a mixture of the composite degrees of freedom $\tilde M_1, \tilde M_2$ and the tadpole is generated by the monopole condensate. 

Let us examine this more carefully. Setting $F_{\tilde M_3, T, Y, Z} = 0$ gives 
\begin{equation}
\label{e.model1otherVEV}
\langle \tilde M_3 \rangle = \langle T \rangle = \langle Z \rangle = 0 \ \ , \ \ \ \ \ \langle Y \rangle = - c \frac{\langle E_+ \tilde E_+\rangle }{m_Y}.
\end{equation}
To ensure that we have massless monopoles, set $F_{\tilde E_+, E_+} = 0$ by enforcing
\begin{equation}
\langle \tilde M_1 \rangle = - \frac{b}{a} \langle \tilde M_2 \rangle~.
\label{M1vev}
\end{equation}
If the monopoles condense the remainder of the theory looks exactly like the Shih-O'Raifeartaigh model. Minimizing $|F_{\tilde M_1}|^2 + |F_{\tilde M_2}|^2$ under the assumption that $\phi_i = 0$ gives the monopole VEV
\begin{equation}
\langle E_+ \tilde E_+ \rangle = \frac{a}{a^2 + b^2} \mu^2.
\end{equation}
The $\tilde M_1$ tadpole ensures that monopole condensation is energetically preferable. The tree-level vacuum energy is
\begin{equation}
\langle V_0 \rangle = \frac{b^2}{a^2 + b^2} \mu^4,
\end{equation}
and receives contributions from both nonzero $F_{\tilde M_{1,2}}$.

It is now clear that in the low-energy limit this theory resembles the Shih-O'Raifeartaigh model, with the pseudomodulus $X$ corresponding to a mixture of the composite $\tilde M_1$ and $\tilde M_2$ while the tadpole $f X$ is generated by monopole condensation, with $f \sim \mu^2$.
(In fact we may simply set $X = \tilde M_2$ and $f =  a b \mu^2/(a^2+b^2) \sim m \Lambda$. The $\tilde M_1$ content of the pseudomodulus has no effect other than to rescale its mass by an $\mathcal{O}(1)$ factor with respect to the corresponding expression for the Shih-O'Raifeartaigh model.)  

To summarize, we have shown that the point
\begin{eqnarray}
\nonumber &\langle \tilde M_3 \rangle = \langle T \rangle = \langle Z \rangle = \langle \phi_i \rangle = 0, \ \ \ \ \ \
\langle Y \rangle = - \frac{c \langle E_+ \tilde E_+\rangle}{m_Y}, \ \ \ \ \  \  \ \langle E_+ \tilde E_+ \rangle = \frac{a}{a^2 + b^2} \mu^2,&\\ \label{e.CEFS1vevsummary}
&\langle \tilde M_1 \rangle = -\frac{b}{a} X, \ \ \ \ \ \ \ \langle \tilde M_2 \rangle = X,&
\end{eqnarray}
constitutes a tree-level stable pseudomoduli space parameterized by the value of $\tilde M_2$. The tree-level spectrum can be divided into four groups:
\begin{itemize} \itemsep=1mm
\item $\tilde M_1, \tilde M_2, \tilde M_3$: These three chiral superfields have a supersymmetric spectrum. There are two massive modes with masses $\mathcal{O}(m_Y, \mu)$ and one zero mode chiral superfield. The fermion component of the zero mode is the Goldstino. The complex scalar component of the zero mode multiplet is $X$. $|X|$ is the pseudomodulus and receives a VEV at 1-loop level, whereas the phase of $X$ is the Goldstone boson of the global $U(1)_R$ under which $X$ has charge $+2$. This is not a global symmetry of the electric superpotential, but is an accidental symmetry in the IR when irrelevant (nonrenormalizable) interactions are neglected.
\item $T, Z$: Their spectrum is also supersymmetric and massive with masses $m_Z + \mathcal{O}(\mu \sqrt{m_Z/\Lambda})$.
\item $Y, E_+, \tilde E_+$: Two massive chiral superfields have the same mass as the non-zero modes in the $\tilde M_i$ group. The other superfield is eaten by the magnetic gauge superfield since the $U(1)_{mag}$ is broken by the monopole VEV.
\item $\phi_1, \phi_2, \phi_3$: The scalar and fermion masses of these fields are identical to the corresponding masses from the Shih-O'Raifeartaigh model (with the substitution  $f \rightarrow a b \mu^2/(a^2 + b^2)$), and are furthermore the only masses that depend on the pseudomodulus. 
\end{itemize}
The 1-loop Coleman-Weinberg potential for $X$ is generated exclusively by the $\phi_i$ masses, giving us an effective low-energy 
Shih-O'Raifeartaigh model below the monopole condensation scale and a corresponding SUSY-breaking vacuum. All the results from 
Section \ref{s.shih} carry over and apply near the origin of our field perturbations.

\subsection{Vacuum stability vs.  incalculable corrections}
\label{ss.model1scaleconstraints}

We will now check what constraints the various scales in the theory must satisfy to ensure that the Shih-O'Raifeartaigh metastable SUSY-breaking vacuum of the deformed $SU(2)^3$ model is not wiped out by $1/\Lambda$ suppressed corrections which we have so far neglected under the assumption that they would be small in the neighborhood of the monopole singularity. 

There are two sources for these corrections: (a) irrelevant  terms in the superpotential, Eqns. (\ref{e.CEFSsuperpot1}) \& (\ref{e.CEFSdeformation1}), and (b)  cubic and higher order terms in the fields in the \kahler potential. We can ignore the higher-order corrections in evaluating the Coleman-Weinberg potential, since their contributions are subdominant to the tree-level mass-dependence on the pseudomodulus $|X|$ (this  will  be demonstrated below). That means we must check two things:  that the higher order corrections do not destabilize any field directions that were flat prior to taking those corrections into account, and that those corrections do not overpower the Coleman-Weinberg potential and destabilize the $|X|$ VEV. 

Since the flat direction corresponding to the Goldstone boson of the broken $U(1)$ is protected, and assuming that the tree-level masses of $M_3$, $T$, $Z$, $Y$, $\phi_1$, and $\phi_2$ are sufficiently large, all we need to worry about are the 
$1/\Lambda$ suppressed corrections involving the pseudo modulus $X$.  The dangerous ones are  \kahler terms cubic in $X$ and non-renormalizable superpotential terms.  
These terms are allowed because of the spontaneous breaking (\ref{M1vev}) and explicit breaking (\ref{e.model1Wtree}) of the $\mathbb{Z}_4$ global symmetry to $\mathbb{Z}_2$.  Both types contribute terms of the form
\begin{equation}
\label{e.CEFS1HOT12}
\delta V \sim V_0 \frac{X + X^\dagger}{\Lambda} \sim m^2 \Lambda (X + X^\dagger)~.
\end{equation}

The potential for $X$ generated by $V_\mathrm{CW}$ looks like a mexican hat in the $X$-complex-plane. The phase is undetermined, but $|X|$ receives a VEV from $V_\mathrm{CW}$. Adding terms like \eref{CEFS1HOT12}, i.e. linear terms in $X$, will generate a definite VEV for $\theta_X$ while shifting the VEV of $|X|$. So to ensure stability, we must check that the linear tilt due to a deformation like $\delta V \sim m^2 \Lambda X$ does not overpower the stabilizing effect of $V_\mathrm{CW}(|X|)$. (We now drop the absolute value signs and use $X$ to describe the component of that field along the tilt direction.) The potential for $X$ including higher order corrections looks schematically like this:
\begin{equation}
V(X) \sim V_\mathrm{CW}(X) + m^2 \Lambda X
\end{equation}
(Note that we have replaced $\mu^2$ by $m \Lambda$, which is sufficient for the required order-of-magnitude estimates.)
To make sure that the local minimum of $V(X)$ is not destroyed by the tilt, we require that the rough scale of the gradient of the potential barrier is much larger than the scale of the gradient of the tilt. Using the result of our numerical scan in \eref{shihbarriergradient}, we obtain the following inequality which must be satisfied to ensure that our mechanism of SUSY-breaking survives the effect of higher-order corrections:
\begin{equation}
\label{e.CEFS1HOT12constraint}
\left(\frac{\Delta V}{\Delta X}\right)_{max}  \gg  m^2 \Lambda \ \ \ \ \ \ \ \Rightarrow \ \ \ \ \ \ 
\mathcal{O}(10^{-2}) \left( \frac{\Lambda}{\Lambda_{UV}}\right)^3 \frac{m^2 \Lambda^2}{m_2} \gg  m^2 \Lambda~,
\end{equation}
which can be simplified to
\begin{equation}
\label{e.model1constraint1}
\frac{m_2}{\Lambda} \  \ll \  \mathcal{O}(10^{-2}) \times \left( \frac{\Lambda}{\Lambda_{UV}}\right)^3.
\end{equation}
There is also another constraint on the scales from SUSY-breaking: 
\begin{equation}
\frac{\lambda m \Lambda}{m_1 m_2} \sim y  < y_{max} \sim \frac{1}{r},
\end{equation}
which becomes
\begin{equation}
\label{e.model1constraint2}
\frac{\Lambda}{\Lambda_{UV}} \lsim \left( \frac{m_1}{\Lambda}\right)^2 \left(\frac{\Lambda}{m}\right)
\end{equation}

To illustrate these constraints, define the powers $c_{UV}, c_m, c_1, c_2$ such that
\begin{equation}
\frac{\Lambda}{\Lambda_{UV}} \sim 10^{-c_{UV}} \ \ , \ \ \ \ \ \ 
\frac{m}{\Lambda} \sim 10^{-c_{m}} \ \ , \ \ \ \ \ \ 
\frac{m_1}{\Lambda} \sim 10^{-c_1}  \ \ , \ \ \ \ \ \ 
\frac{m_2}{\Lambda} \sim 10^{-c_2} \ \ .
\end{equation}
Then equations (\ref{e.model1constraint1}), (\ref{e.model1constraint2})   imply
\begin{equation}
\label{e.model1constraints}
c_2 > 2 + 3 c_{UV} \ \  , \ \ \ \ \ \ \ \ c_1 \lsim \frac{1}{2} (c_m + c_{UV}),
\end{equation}
in addition to  $c_{UV}, c_m \geq 2$ which protects the nonperturbative monopole physics. There are clearly a variety of ways that this can be satisfied. For example, we could reasonably set $\Lambda/\Lambda_{UV} \sim 0.01$ and $m_1 \sim m_2$, i.e. $c_{UV} = 2, c_1 = c_2$. Then $c_m \geq14$ and $c_{1,2} = 8$ when the inequality is saturated. (For $c_m > 14$, $c_{1,2}$ can be somewhat larger.) In this case the hierarchies of the model are
\begin{equation}
m \ll m_{1,2} \ll \Lambda \ll \Lambda_{UV}.
\end{equation}

Finally, to ensure that $\tilde M_3$ is not destabilized by \kahler corrections $m_Y \sim a_Y \Lambda \ll \Lambda$ cannot be too small. The lower bound is
\begin{equation}
\label{e.model1ayconstraint}
a_Y \gg \frac{m}{\Lambda}.
\end{equation}

We emphasize that these constraints, while restricting us to certain areas of the model's parameter space, do not represent tuning. There is no particular balancing of parameters involved in stabilizing the false vacuum. The above hierarchies merely guarantee that certain potentially destabilizing contributions are subdominant. 

\subsection{Weak-coupling limit and supersymmetric runaways}

By inspection of \eref{model1Wtree} it is clear that in the weak coupling or classical limit ($\Lambda \rightarrow 0$), supersymmetry is restored with one supersymmetric vacuum at the origin of field space: $Q_i = \phi_i = Z = Y = 0$. This means that supersymmetry breaking depends on the nonperturbative $SU(2)^3$ dynamics.

In the $\Lambda \neq 0$ case this model has two runaways which both resemble the runaway in
the  Shih-O'Raifeartaigh model. The first runaway\footnote{
Using the word runaway implies that there is a SUSY-breaking minimum along this runaway at $X = \infty$, but it is in fact more likely that the fields would eventually roll into the SUSY-runaway. 
} takes $F_{\tilde M_2}, F_{\phi_i} \rightarrow 0$ in a manner analogous to \eref{shihrunaway}, but this is not a supersymmetric runaway since $F_{\tilde M_1} \neq 0$.  The other runaway is the only supersymmetric runaway in this model. Increasing the monopole VEV from $\langle E_+ \tilde  E_+\rangle =  a \mu^2/(a^2+b^2)$ to $\mu^2/a$ sets $F_{\tilde M_1} = 0$. $F_{\tilde M_2}, F_{\phi_i}$ are identical to the $F$-terms in the Shih-O'Raifeartaigh model with the replacement $X \rightarrow \tilde M_2$ and $f \rightarrow \frac{b \mu^2}{a}$, and are taken to zero via the Shih-O'Raifeartaigh runaway, \eref{shihrunaway}. In both cases $F_{Y,Z,T,\tilde M_3} = 0$ via the VEVs in \eref{model1otherVEV} just like in the SUSY-breaking minimum, and $\tilde M_1$ is free to move however it has to in order to set $F_{E, \tilde E} = 0$. The trajectory of $\tilde M_1$ depends on the other fields and implicitly includes all the  corrections to the monopole mass in \eref{CEFSsuperpot1} that we can ignore close to the origin of our perturbations.

Assuming we can trust our approximately canonical \kahler potential, the potential energy along \emph{both} runaways
\begin{equation}
V_{\cancel{SUSY}\ run} = \frac{b^4 \mu^4}{(a^2+b^2)^2} + \frac{m_1^2 m_2 \mu^4}{\lambda^2 |\tilde M_2|} \frac{a b}{a^2 + b^2} \ \ , \ \ \ \ 
V_{SUSY\ run} = \frac{m_1^2 m_2 \mu^2}{\lambda^2 |\tilde M_2|} \frac{b}{a}
\end{equation}
becomes equal to the potential energy of the SUSY-breaking pseudomoduli space
\begin{equation}
V_{PMS} = \frac{b^2}{a^2+b^2} \mu^4.
\end{equation}
when
\begin{equation}
{\tilde M_2} = \frac{m_1^2 m_2}{\lambda^2 \mu^2} \frac{a^2 + b^2}{a b}.
\end{equation}
For the explicit examples of scales considered below \eref{model1constraints}, this is much less than $\Lambda$, making the calculation trustworthy, but even if noncanonical \kahler corrections become important it would not significantly change the result that the \emph{width} of the potential barrier between the SUSY-breaking pseudomoduli space and both runaways is of roughly the same size along the $X$ direction. (The same can be said for the distance along the $\phi_i$ directions, since along both runaways $\phi_i$ scales with $\langle E_+ \tilde E_+ \rangle^{1/2} \sim \mu$.) However, since the SUSY-runaway is separated from the SUSY-breaking local minimum by an $\mathcal{O}(\mu^4)$ potential barrier, the decay path of the false vacuum is much more likely to be across the tiny barrier created by $V_\mathrm{CW}$ to either the SUSY-breaking or supersymmetric runaway. Therefore, since $\lambda$ is small, the same arguments that ensure longevity of the false vacuum in the Shih-O'Raifeartaigh model apply here as well.

\subsection{Aligning the electric deformations}

There is one possible source of tuning in this model, which is the alignment of the deformations in the electric theory. If the coefficients of $Q_i^2$ and $\Lambda^2$ in the linear combinations $(QQ)_{A,B,C}$ are not properly chosen in the electric superpotential \eref{model1Wtree} then they do not correspond to the canonical IR degrees of freedom $\tilde M_i$ and the effective IR superpotential will not exactly resemble \eref{model1deformations}. 

We can get a feeling for the required level of alignment by considering the following more general superpotential,
\begin{eqnarray}
\nonumber \delta W &=& - \mu^2 (\tilde M_1 + \epsilon_{12} \tilde M_2 + \epsilon_{13} \tilde M_3) + \lambda (\tilde M_2 + \epsilon_{21} \tilde M_1 + \epsilon_{23} \tilde M_3) \phi_1 \phi_2 \\
&&+ \frac{m_2}{2} \phi_2^2 + m_1 \phi_1 \phi_3
 + m_Z Z T + m_Y (\tilde M_3  + \epsilon_{31} \tilde M_1 + \epsilon_{32} \tilde M_2)Y,
\end{eqnarray}
Most of these misalignments are harmless, shifting tree-level VEVs or inducing tree-level tadpoles for the pseudomodulus. However, some can destabilize the SUSY-breaking vacuum.

The $\epsilon_{21}$ term, apart from inducing a tree-level pseudomodulus tadpole like $\epsilon_{12}$, shifts the effective $\lambda$-coupling in the fermion contribution to $V_\mathrm{CW}$.  For $r < 10$, the maximum allowed values of $\epsilon_{21}$ that do not destroy the local minimum lie in the range $\epsilon_{21}^{max} \sim 10^{-2} - 10^{-7}$ depending on $r$ and $y$, with $\epsilon_{21}^{max}$ decreasing for larger $r$ and smaller $y$. This represents the required level of tuning in the $(QQ)_{A,B}$ linear combinations.

The $\epsilon_{31}$ and $\epsilon_{32}$ terms give a mass to  the pseudomodulus  $m_X = \epsilon_Y m_Y \sim \epsilon_Y a_Y \Lambda$, where $\epsilon_Y = \frac{1}{2} (\epsilon_{32} - \frac{b}{a} \epsilon_{31})$. Adding a pseudomodulus mass to the Shih-O'Raifeartaigh model creates a SUSY minimum at $X = f/m_X$ and tilts the pseudomoduli space away from the origin. We have to make sure that $V_\mathrm{CW}$ is not overwhelmed by this gradient, which at a position $\langle X \rangle$ on the pseudomoduli space is given by
\begin{equation} 
\left( \frac{\Delta V}{\Delta X}\right)_{mX} = m_X^2 \langle X \rangle + f m_X.
\end{equation}
Translating this to our effective Shih-O'Raifeartaigh model and using Equations (\ref{e.shihXvev}), (\ref{e.shihbarriergradient}), we obtain the following upper bounds:
\begin{equation}
\label{e.epsilonYtuning}
\epsilon_Y a_Y \ll \left\{ \begin{array}{ll} \frac{1}{8 \pi^2} \left( \frac{\Lambda}{\Lambda_{UV}}\right)^3 \frac{m}{m_2} & \mbox{when $r \lsim 2$, i.e. $\langle X \rangle = 0$}\\
\frac{1}{\sqrt{8 \pi^2}} \left(\frac{\Lambda}{\Lambda_{UV}}\right)^2 \frac{m}{m_2}
& \mbox{when $r \gsim 2$, i.e. $\langle X \rangle \neq 0$}
\end{array}\right.,
\end{equation}
where we have used the fact that for small $\lambda$, $\langle X \rangle \gg f$ when $r \gsim 2$.

Consider the explicit example of scales considered below equation \eref{model1constraints}. $a_Y$ is already a small number satisfying $10^{-14} \ll a_Y \ll 1$. The constraint \eref{epsilonYtuning} gives
 \begin{equation}
 \epsilon_Y a_Y \ll \left\{ \begin{array}{ll} 
 \sim 10^{-16}
  & \mbox{when $r \lsim 2$, i.e. $\langle X \rangle = 0$}\\
\sim 10^{-13}
 & \mbox{when $r \gsim 2$, i.e. $\langle X \rangle \neq 0$}
\end{array}\right..
\end{equation}
If we take $a_Y \sim 10^{-13}$, $\epsilon_Y$ could be as big as $10^{-3}$ for $\langle X \rangle = 0$ and unity for $\langle X \rangle \neq 0$, representing the required level of tuning in the $(QQ)_C$ linear combination.

\section{Other SUSY-breaking deformations of the $SU(2)^3$ model}
\label{s.model2and3} \setcounter{equation}{0} \setcounter{footnote}{0}

In \sref{model1} we constructed a theory of monopole-triggered SUSY-breaking based on the $SU(2)^3$ model. This method of deforming models with massless monopoles to induce SUSY-breaking seems fairly general, and it is instructive to try and embed the Shih-model differently into the monopole sector. 

The first alternative possibility is to make $\phi_2$ composite instead of the pseudomodulus $X$. Starting from \eref{CEFSsuperpot1}, this would lead us to add the deformations and singlet fields
\begin{equation}
\label{e.model2deformations}
\delta W = - f \tilde M_1 + X (\lambda T \phi_1 - \mu^2) + m_1 \phi_1 \phi_3 + m_Z Z \tilde M_2 + m_Y Y \tilde M_3.
\end{equation}
These couplings preserve the $\mathbb{Z}_2$ global symmetry as long as $\phi_1$ and $ \phi_3$ are also charged. The $\tilde M_1$ tadpole induces a monopole VEV which generates a mass for $T$, completing the Shih-O'Raifeartaigh sector. The coupling $\lambda \sim (\Lambda/\Lambda_{UV})^2$ comes from an operator that is higher dimensional than in the model of Section \ref{s.model1}. The stability of $X$ against incalculable \kahler corrections as well as the existence of a SUSY-breaking minimum requires the hierarchy
\begin{equation}
m_1 \ll \mu \ \ ,  \ \ \ \ \ \ m \ll \Lambda \ll \Lambda_{UV}.
\end{equation}
Turning off the $SU(2)^3$ gauge interactions ($\Lambda \rightarrow 0$) does not destroy the SUSY-breaking pseudomoduli space at tree-level, meaning the gauge interactions deform a classically existing SUSY-breaking minimum, similar to \cite{AMOS}. Since the goal of this paper is to find a model of dynamical monopole-triggered SUSY-breaking we will not pursue this possibility further. 

The `most dynamical' embedding  of the Shih-O'Raifeartaigh sector into the $SU(2)^3$  model is to make both the pseudomodulus $X$ and $\phi_2$ composite. Since we are using `more' of the monopole sector to break SUSY this requires fewer deformations:
\begin{equation}
\label{e.model3deformations}
\delta W = - f \tilde M_1 + \lambda \tilde M_2 T \phi_1 + m_1 \phi_1 \phi_3 + m_Y Y M_3.
\end{equation}
Again the $\tilde M_1$ tadpole induces a monopole VEV, which now provides a tadpole for $\tilde M_2$ as well as a mass for $T$, which act as the pseudomodulus and $\phi_2$ respectively. Stability of $X$ against \kahler corrections and existence of the SUSY-breaking vacuum requires $m \ll \Lambda$ due to the smallness of $\lambda \sim (\Lambda/\Lambda_{UV})^3$:
\begin{equation}
\left(\frac{m}{\Lambda}\right) \ll \left( \frac{\Lambda}{\Lambda_{UV}}\right)^9 \ \ , \ \ \ \ \ 
\frac{m_1}{\Lambda} \gsim \left(\frac{\Lambda}{\Lambda_{UV}}\right)^3.
\end{equation}
As desired, SUSY is restored in the $\Lambda \rightarrow 0$ limit. Both of these models feature standard Shih-O'Raifeartaigh runaways in the strong coupling case, with $\tilde M_1$ adjusting to keep the trajectory on the singularity. 

This latter model appears more elegant than the model of \sref{model1}, but it suffers the unfortunate drawback that \kahler corrections render the $T$-mass incalculable, a result of the monopole VEV doing double duty. While this also does not fulfill our requirements for a calculable monopole-triggered dynamical SUSY-breaking theory, these two alternative deformations of the $SU(2)^3$ model demonstrate how one might produce many more models of SUSY-breaking that include monopole dynamics.

\section{Conclusions}
\label{s.conclusion} \setcounter{equation}{0} \setcounter{footnote}{0}

Monopoles have many unique characteristics that make them very interesting. Their unusual dynamics might hold the key to constructing novel models of supersymmetry (or perhaps electroweak) breaking. Topological monopoles, traditionally treated as nonperturbative objects, can be calculationally controlled using Seiberg-Witten methods. This opens up new avenues for model building. 

In our model, supersymmetry breaking is triggered by monopole condensation. A suitably deformed $SU(2)^3$ theory with massless monopoles takes on the form of an effective Shih-O'Raifeartaigh model with a meta-stable SUSY-breaking local minimum. In constructing such a model within the limitations of $\mathcal{N} = 1$ supersymmetry it is important to check that incalculable \kahler corrections do not destabilize the false vacuum. We have shown that these contributions can be controlled, and through appropriate choices of scales can be made subdominant. Additional deformations of the $SU(2)^3$ model, with varying characteristics, demonstrate more generally how models with massless monopoles might be deformed to induce SUSY-breaking. It is our hope that this will open up new investigations which might eventually yield elegant mechanisms of breaking supersymmetry that circumvent some of the problems encountered by other approaches. It would be interesting to explore SUSY-breaking in theories that have both massless electrically and magnetically charged particles.

\subsection*{Acknowledgements}
We would like to thank Matt Baumgart, Luis \'Alvarez-Gaum\'e, Mario Martone and Zohar Komargodski for 
valuable conversations. 
C.C., Y.S. and J.T thank the Aspen Center for Physics and C.C. and J.T. also the Kavli Institute for Theoretical Physics where part of this work was completed.

The work of C.C. and D.C. was supported in part by the National Science Foundation under grant PHY-0757868.  V.R. was supported by Department of Energy under grant DE-FG02-04ER-41298. Y.S. was supported by National Science Foundation under grant PHY-0970173. J.T. was supported by the Department of Energy under grant
DE-FG02-91ER406746.

\end{document}